\documentclass[9pt,english,conference]{IEEEtran}
\usepackage[T1]{fontenc}
\usepackage[latin1]{inputenc}
\usepackage{geometry}
\usepackage{color}
\usepackage{amsthm}
\usepackage{amsmath}
\usepackage{amssymb}
\usepackage{graphicx}
\usepackage{setspace}
\usepackage{esint}
\makeatletter
\theoremstyle{plain}
\newtheorem{thm}{\protect\theoremname}
\theoremstyle{definition}
\newtheorem{defn}{\protect\definitionname}
\theoremstyle{remark}
\newtheorem{rem}{\protect\remarkname}

\usepackage{babel}
\usepackage{url}

\usepackage{babel}

\usepackage{babel}
\providecommand{\theoremname}{Theorem}

\allowdisplaybreaks

\usepackage{babel}

\providecommand{\definitionname}{Definition}

\providecommand{\theoremname}{Theorem}

\usepackage{babel}

\providecommand{\definitionname}{Definition}
\providecommand{\theoremname}{Theorem}

\usepackage{babel}

\providecommand{\definitionname}{Definition}
\providecommand{\remarkname}{Remark}
\providecommand{\theoremname}{Theorem}

\usepackage{babel}

\providecommand{\definitionname}{\inputencoding{latin9}Definition}
\providecommand{\remarkname}{\inputencoding{latin9}Remark}
\providecommand{\theoremname}{\inputencoding{latin9}Theorem}

\usepackage{babel}
\providecommand{\definitionname}{Definition}
\providecommand{\remarkname}{Remark}
\providecommand{\theoremname}{Theorem}

\usepackage{babel}
\providecommand{\definitionname}{Definition}
\providecommand{\remarkname}{Remark}
\providecommand{\theoremname}{Theorem}

\makeatother

\usepackage{babel}
\providecommand{\definitionname}{Definition}
\providecommand{\remarkname}{Remark}
\providecommand{\theoremname}{Theorem}

\begin{document}

\title{On the Ice-Wine Problem: Recovering Linear Combination of Codewords
over the Gaussian Multiple Access Channel}

\author{\authorblockA{Shahab Ghasemi-Goojani and Hamid Behroozi\\
 Department of Electrical Engineering, Sharif University of Technology,
Tehran, Iran\\
 Email: ghasemeishahab@gmail.com, behroozi@sharif.edu}\vspace{-6mm}
 }
\maketitle
\begin{abstract}
In this paper, we consider the Ice-Wine problem: Two transmitters
send their messages over the Gaussian Multiple-Access Channel (MAC)
and a receiver aims to recover a linear combination of codewords.
The best known achievable rate-region for this problem is due to \cite{Nazar_IT_11,Wilson_IT_2010_Relay}
as $R_{i}\leq\frac{1}{2}\log\left(\frac{1}{2}+{\rm SNR}\right)$ $(i=1,2)$.
In this paper, we design a novel scheme using lattice codes and show
that the rate region of this problem can be improved. The main difference
between our proposed scheme with known schemes in \cite{Nazar_IT_11,Wilson_IT_2010_Relay}
is that instead of recovering the sum of codewords at the decoder,
a non-integer linear combination of codewords is recovered. Comparing
the achievable rate-region with the outer bound, $R_{i}\leq\frac{1}{2}\log\left(1+{\rm SNR}\right)\,\,(i=1,2)$,
we observe that the achievable rate for each user is partially tight.
Finally, by applying our proposed scheme to the Gaussian Two Way Relay
Channel (GTWRC), we show that the best rate region for this problem
can be improved.
\end{abstract}

\section{\label{sec:Introduction}Introduction}

Lattice structures have been shown to be capacity-achieving for AWGN
channels such as the Gaussian point-to-point channel \cite{Erez_IT_04},
Multiple Access Channel (MAC) \cite{Nazar_IT_11}, Broadcast Channel
(BC) \cite{Zamir_02} and relay networks \cite{Nazar_IT_11}. Nested
lattice codes have been shown to achieve the same rates which are
achievable by independent, identically distributed (i.i.d) \textcolor{black}{Gaussian}
random codes in the decode-and-forward and compress-and-forward schemes
for the relay channel \cite{Song_IT_13}. However, in some scenarios,\textcolor{black}{{}
}lattice codes may outperform i.i.d. random codes particularly when
we are interested in decoding a linear combination of codewords rather
than decoding the individual codewords as the compute-and-forward
scheme \cite{Nazar_IT_11}.

\begin{figure}
\begin{centering}
\includegraphics[width=8cm]{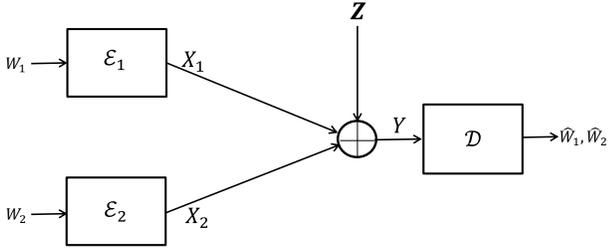} 
\par\end{centering}

\caption{\label{fig:Fig1}The Gaussian Multiple Access Channel (MAC).}
\end{figure}

The compute-and-forward scheme \cite{Nazar_IT_11} is a novel strategy
which uses the advantage of the linear structure in lattice codes
and the additive nature of Gaussian networks in order to get some
new achievable rate-regions for decoding linear combination of messages.
Consider the multiple access communication system model depicted in
Fig. \ref{fig:Fig1}, which can be seen such as a basic element for
the relay networks. Each sender wishes to communicate an independent
message reliably to a common receiver. In \cite{Ahlswede_ISIT_71},
it is shown that the capacity region of the Gaussian MAC is given
by the following rate region:
\begin{eqnarray*}
R_{i} & \leq & \frac{1}{2}\log\left(1+\frac{P}{N}\right),\,\,(i=1,2)\\
R_{1}+R_{2} & \leq & \frac{1}{2}\log\left(1+\frac{2P}{N}\right),
\end{eqnarray*}
where $P$ is an average transmit power constraint at both nodes and
$N$ is the noise variance. Now, suppose that instead of estimating
transmitted codewords $\boldsymbol{X}_{1}$ and $\boldsymbol{X}_{2}$
individually, we are interested in decoding the sum of codewords (or
messages), i.e., $\boldsymbol{X}_{1}+\boldsymbol{X}_{2}$. This problem
is called the Ice-Wine problem \cite{Gastpar_ISIT2011_Tut}. One approach
for solving this problem is based on random codes, i.e., codes from
a random ensemble. For this purpose, we must first recover both messages
and then recover the desired function. Since only a function of messages
is desirable (instead of both messages separately), this approach
is not optimal.

For this problem, in \cite{Popovski_07} it is conjectured that a
rate-region of $R_{i}<\frac{1}{2}\log\left(1+{\rm SNR}\right)$ $(i=1,2)$
can be achieved, however, no transmission scheme is provided. A constructive
scheme is proposed independently in \cite{Nazar_IT_11} and \cite{Wilson_IT_2010_Relay}.
In \cite{Wilson_IT_2010_Relay}, to decode the sum of codewords modulo
a lattice, two schemes are proposed: one lattice coding scheme based
on minimum angle decoding while the other (which is similar to the
one used by Nazer and Gastpar \cite{Nazar_IT_11}) is based on the
proposed scheme in \cite{Erez_IT_04} for the AWGN channel. Nazer
and Gastpar used the compute-and-forward scheme to obtain any arbitrary
integer linear combination of messages. They applied this idea to
relay networks to achieve some new rate-regions \cite{Nazar_IT_11}.
In both these papers it is shown that for this problem the best achievable
rate-region is $R_{i}\leq\frac{1}{2}\log\left(\frac{1}{2}+{\rm SNR}\right)$
$(i=1,2)$. As we can see, there is a loss at most 1/2 bit. Recently,
Zhan, Nazer, Erez and Gastpar proposed a new linear receiver architecture,
called Integer-Forcing \cite{Zhan_arxiv_2014}, where the decoder
recovers integer combinations of the codewords. They use the receiver
antennas to create an effective channel matrix with integer-valued
element. Although, there have been some attempts to improve the achievable
rate-region of the compute-and-forward scheme for decoding the sum
of messages \cite{Ordentlich_13_Allerton,Zhan_arxiv_2014}, the authors
in \cite{Zhan_arxiv_2014} show that this scheme is not able to achieve
a larger rate-region than the compute-and-forward scheme.

The compute-and-forward scheme was used in subsequent works to achieve
new rate-regions in many networks, see e.g. \cite{Nam_IT_2010,Nam_IT_2011}.
In \cite{Nam_IT_2010} the compute-and-forward scheme is applied to
the Gaussian Two-Way Relay Channel (GTWRC) to achieve the capacity
region for this channel within 1/2 bit. By modifying the compute-and-forward
for the Gaussian MAC with unequal powers, in \cite{Nam_IT_2011} it
is shown that for the Gaussian relay networks with interference, the
multicast capacity is achievable within a constant gap which depends
on only the number of users. Note that in this class of relay networks,
at each node, outgoing channels to its neighbors are orthogonal, while
incoming signals from neighbors can interfere with each other.\textcolor{black}{{}
More recently, Zhu and Gastpar proposed a modifi{}ed compute-and-forward
scheme that is based on channel state information at the transmitters
(CSIT) in order to compute the linear combination over the Gaussian
MAC \cite{Zhu_IZS_14}. Then, using numerical results, they show that
this scheme can achieve a rate-region that is better than that of
the common compute-and-forward scheme. Also, by applying it to the
GTWRC, they shown that it can improve the best rate-region of the
GTWRC which is obtained in \cite{Nam_IT_2011}.}

In this paper, we use structured lattice codes to obtain a new rate-region
for the Ice-Wine problem. In all previous attempts, the sum of codewords
is decoded and it is shown that there is a gap between the achievable
rate and the upper bound for any finite SNR. This paper aims to answer
the open challenge of getting the full ``one plus\textquotedblright{}
term in the achievable rate of each user. Although reaching this goal
does not seem to be feasible with nested lattices, in this paper,
using nested lattice codes, we decode a non-integer linear combination
of codewords, $\boldsymbol{V}_{1}+\alpha\boldsymbol{V}_{2}$, instead
of an integer linear combination of codewords. For this purpose, we
first construct a lattice chain at the transmitter where the codebook
at one transmitter depends on $\alpha$. As we will see, we can achieve
the full rate for one user but due to the chosen codebooks, we can
not achieve the full rate for the other user.\textcolor{red}{{} }\textcolor{black}{Although
we were not aware of this recent work of Zhu and Gastpar \cite{Zhu_IZS_14}
at the time we submitted this paper to Information Theory Workshop
(ITW) 2014, but the main difference between our proposed scheme and
the new scheme of Zhu and Gatspar is due to the fact that in the scheme
of \cite{Zhu_IZS_14}, we must set the CSIT such that the achievable
rate is maximized. But in our proposed scheme, we try to decrease
the variance of the effective noise which helps us to get a rate that
is better than that of the common compute-and-forward scheme. This
distinguishes our proposed scheme with that of \cite{Zhu_IZS_14}.}
As an application of our proposed scheme, we apply it to the GTWRC
and we show that the best rate-region given in \cite{Nam_IT_2010}
for this open problem can be improved.

The remainder of the paper is organized as follows. Section \ref{sec:Lattice-Codes}
provides a brief review of nested lattice codes. In Section \ref{sec:Our-Proposed-Scheme},
we present our proposed scheme for the Ice-Wine problem. Section \ref{sec:Conclusion}
concludes the paper.

\section{\label{sec:Lattice-Codes}Lattice Codes}

Here, we provide some necessary definitions on lattices and nested
lattice codes. Interested reader can refer to \cite{Nazar_IT_11,Erez_IT_04,conway_book}
and the references therein for more details. 
\begin{defn}
A lattice $\Lambda$ is a discrete additive subgroup of $\mathbb{R}^{n}$.
A lattice $\Lambda$ can always be written in terms of a generator
matrix $\mathbf{G}\in\mathbb{R}^{n\times n}$ as $\Lambda=\left\{ \boldsymbol{x}=\boldsymbol{z}\mathbf{G}:\boldsymbol{z}\in\mathbb{Z}^{n}\right\} ,$
where $\mathbb{Z}$ represents the set of integers.

The \textit{\textcolor{black}{nearest neighbor quantizer}} $\mathcal{Q}_{\Lambda}$
maps any point $\boldsymbol{x}\in\mathbb{R}^{n}$ to the nearest lattice
point: 
\[
\mathcal{Q}_{\Lambda}(\boldsymbol{x})=\arg\underset{\boldsymbol{l}\in\Lambda}{\min}\left\Vert \boldsymbol{x}-\boldsymbol{l}\right\Vert .
\]
The \textit{\textcolor{black}{fundamental Voronoi region}} of lattice
$\Lambda$ is set of points in $\mathbb{R}^{n}$ closest to the zero
codeword, i.e., 
\[
\mathcal{V}_{0}(\Lambda)=\left\{ \boldsymbol{x}\in\mathbb{\mathbb{R}}^{n}:\mathcal{Q}(\boldsymbol{x})=0\right\} .
\]
$\sigma^{2}\left(\Lambda\right)$ which is called the second moment
of lattice $\Lambda$ is defined as 
\begin{equation}
\sigma^{2}(\Lambda)=\frac{1}{n}\frac{\int_{\mathcal{V}(\Lambda)}\left\Vert \boldsymbol{x}\right\Vert ^{2}d\boldsymbol{x}}{\int_{\mathcal{V}(\Lambda)}d\boldsymbol{x}},\label{eq:SM}
\end{equation}
and the \textit{\textcolor{black}{normalized second moment}} of lattice
$\Lambda$ can be expressed as 
\[
G(\Lambda)=\frac{\sigma^{2}(\Lambda)}{[\int_{\mathcal{V}(\Lambda)}d\boldsymbol{x}]^{\frac{2}{n}}}=\frac{\sigma^{2}(\Lambda)}{V^{\frac{2}{n}}},
\]
where $V=\int_{\mathcal{V}(\Lambda)}d\boldsymbol{x}$ is the Voronoi
region volume.

The \textit{\textcolor{black}{modulo-}}$\Lambda$ \textit{\textcolor{black}{operation}}
with respect to lattice $\Lambda$ returns the quantization error
\[
\boldsymbol{x}\mbox{ mod }\Lambda=\boldsymbol{x}-\mathcal{Q}(\boldsymbol{x}),
\]
that maps $\boldsymbol{x}$ into a point in the fundamental Voronoi
region and it is always placed in $\mathcal{V}$. The modulo lattice
operation satisfies the following distributive property \cite{Nazar_IT_07}
\[
\left[\boldsymbol{x}\mbox{ mod }\Lambda+\boldsymbol{y}\right]\mbox{ mod }\Lambda=\left[\boldsymbol{x}+\boldsymbol{y}\right]\mbox{ mod }\Lambda.
\]

(Quantization Goodness or Rogers-good): A sequence of lattices $\Lambda\subseteq\mathbb{R}^{n}$
is good for mean-squared error (MSE) quantization if 
\[
\underset{n\rightarrow\infty}{\lim}G\left(\Lambda\right)=\frac{1}{2\pi e}.
\]

\end{defn}
The sequence is indexed by the lattice dimension $n$. The existence
of such lattices is shown in \cite{Zamir_IT_96,Erez_IT_05}. 
\begin{defn}
\label{(AWGN-channel-coding} (AWGN channel coding goodness or Poltyrev-good):
Let $\boldsymbol{Z}$ be a length-$i.i.d$ Gaussian vector, $\boldsymbol{Z}\thicksim\mathcal{N}\left(\boldsymbol{0},\sigma_{Z}^{2}\boldsymbol{I}_{n}\right)$.
The volume-to-noise ratio of a lattice is given by 
\[
\mu\left(\Lambda,\epsilon\right)=\frac{\left(\mbox{ Vol}(\mathcal{V})\right)^{2/n}}{2\pi e\sigma_{Z}^{2}},
\]
where $\sigma_{Z}^{2}$ is chosen such that $\mbox{ Pr}\left\{ \boldsymbol{Z}\notin\mathcal{V}\right\} =\epsilon$
and $\boldsymbol{I}_{n}$ is an $n\times n$ identity matrix. A sequence
of lattices is $\Lambda$ Poltyrev-good if 
\[
\underset{n\rightarrow\infty}{\lim}\mu\left(\Lambda,\epsilon\right)=1,\,\,\,\,\,\forall\epsilon\in\left(0,1\right)
\]
and, for fixed volume-to-noise ratio greater than $1$, $\mbox{ Pr}\left\{ \boldsymbol{Z}\notin\mathcal{V}\right\} $
decays exponentially in $n$ .

(Nested Lattices): A lattice $\Lambda$ is said to be nested in lattice
$\Lambda_{1}$ if $\Lambda\subseteq\Lambda_{1}$. $\Lambda$ is referred
to as the coarse lattice and $\Lambda_{1}$ as the fine lattice.

(Nested Lattice Codes): A nested lattice code is the set of all points
of a fine lattice $\Lambda_{1}$ that are within the fundamental Voronoi
region $\mathcal{V}$ of a coarse lattice $\Lambda$, i.e., $\mathcal{C}=\left\{ \Lambda_{1}\cap\mathcal{V}\right\} .$
The rate of a nested lattice code is defined as 
\[
R=\frac{1}{n}\log\left|\mathcal{C}\right|=\frac{1}{n}\log\frac{\mbox{ Vol}\left(\mathcal{V}\right)}{\mbox{ Vol}\left(\mathcal{V}_{1}\right)}.
\]
In \cite{Erez_IT_05}, Erez, Litsyn and Zamir show that there exists
a sequence of lattices that are simultaneously good for packing, covering,
source coding (Rogers-good), and channel coding (Poltyrev-good).
\end{defn}

\section{\label{sec:Our-Proposed-Scheme}Our Proposed Scheme}

As an achievable scheme, we use a lattice-based coding scheme. In
\cite{Nazar_IT_11,Nam_IT_2011} by using two nested lattice codes,
where one of the lattices provides us codewords while the other lattice
satisfies the power constraint at each user, an achievable rate-region
for the Ice-Wine problem is established. In fact, the decoder recovers
an integer combination of messages. In this paper, we provide a new
achievable rate-region for this problem. To reach this goal, we first
construct three nested lattices where one of them provides codewords
while the other two lattices satisfy the power constraints. At the
destination, instead of finding an integer combination of lattice
points (or messages), we recover a non-integer linear combination
of lattice points. Finally, we apply our proposed scheme to the Gaussian
Two-Way Relay Channel (GTWRC) to improve the best rate-region for
this open problem so far. Let us consider a standard model of a Gaussian
MAC with two users:
\begin{equation}
\boldsymbol{Y}=\boldsymbol{X}_{1}+\boldsymbol{X}_{2}+\boldsymbol{Z},\label{eq:input decoder}
\end{equation}
where $\boldsymbol{Z}$ denotes the AWGN process with zero mean and
variance $N$. Each channel input $\boldsymbol{X}_{i}$ is subject
to an average power constraint $P$, i.e., $\frac{1}{n}\mathbb{E}\left\Vert \boldsymbol{X}_{i}\right\Vert ^{2}\leq P$.

In the following, by applying a lattice-based coding scheme, we obtain
a new achievable rate-region to estimate a linear combination of messages
for the Gaussian MAC. For this purpose, suppose that there exist two
lattices $\Lambda_{1}$ and $\Lambda_{2}$, which are Rogers-good
(i.e.,$\underset{n\rightarrow\infty}{\lim}G\left(\Lambda_{i}^{(n)}\right)=\frac{1}{2\pi e},\,\textrm{for\,\,\,}i=1,2\,)$,
and Poltyrev-good with the following second moments
\[
\sigma^{2}\left(\Lambda_{i}\right)=P.
\]
Also, there is a lattice $\Lambda_{c}$ which is Poltyrev-good with
$\Lambda_{1}\subseteq\alpha_{1}\Lambda_{2}\subseteq\Lambda_{c}$ ($\alpha_{1}$
is a coefficient smaller than one).

\textit{Encoding}: To transmit both messages, we first construct the
following codebooks:
\begin{eqnarray*}
\mathcal{C}_{1} & = & \left\{ \Lambda_{c}\cap\mathcal{V}_{1}\right\} ,\,\,\,\mathcal{C}_{2}=\left\{ \frac{\Lambda_{c}}{\alpha_{1}}\cap\mathcal{V}_{2}\right\} .
\end{eqnarray*}
At each encoder, the message set $\left\{ 1,2,...,2^{nR_{i}}\right\} $
is arbitrarily mapped onto $\mathcal{C}_{i}$ $(i=1,2)$. Then, node
$i$ chooses $\boldsymbol{V}_{i}\in\mathcal{C}_{i}$ associated with
the message $W_{i}$ and sends
\[
\boldsymbol{X}_{i}=\left[\boldsymbol{V}_{i}-\boldsymbol{D}_{i}\right]\textrm{ mod }\Lambda_{i},
\]
where \textcolor{black}{$\boldsymbol{D}_{1}$} and \textcolor{black}{$\boldsymbol{D}_{2}$}
are t\textcolor{black}{wo independent dithers} that are uniformly
distributed over Voronoi regions $\mathcal{V}_{1}$ and $\mathcal{V}_{2}$,
respectively. Dithers are kno\textcolor{black}{wn at the encoders
and the decoder. }Due to the Crypto-lemma \cite{Forney_Allerton_2003},
$\boldsymbol{X}_{i}$ is uniformly distributed over $\mathcal{V}_{i}$
and independent of $\boldsymbol{V}_{i}$. Thus, the average transmit
power of node $i$ equals to $P$, and the power constraint is met.

\begin{figure}
\begin{centering}
\includegraphics[width=9cm]{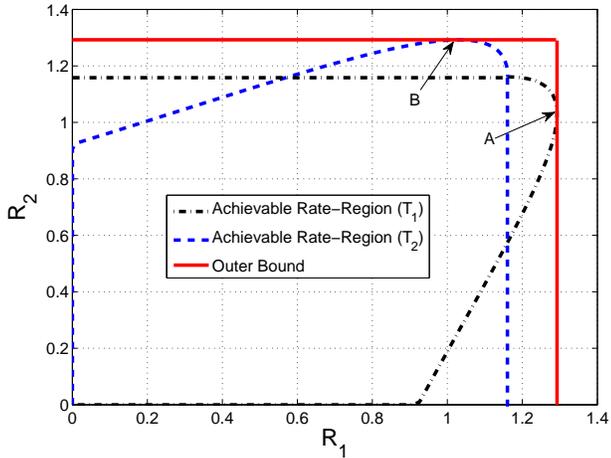}
\par\end{centering}

\caption{\label{fig:Fig2}The achievable rate-region in order to estimate $\boldsymbol{T}_{1}$
and $\boldsymbol{T}_{2}$. As we observe, the proposed scheme can
reach to the outer bound at points $A$ and $B$. The signal-to-noise
ratio is fixed at $SNR=5$. }
\end{figure}

\textit{Decoding}: At the decoder, based on the channel output that
is given by (\ref{eq:input decoder}), we estimate
\[
\left[\boldsymbol{V}_{1}+\alpha_{1}\boldsymbol{V}_{2}-\alpha_{1}\mathcal{Q}_{\Lambda_{2}}\left(\boldsymbol{V}_{2}-\boldsymbol{D}_{2}\right)\right]\textrm{ mod }\Lambda_{1}.
\]
To do this, the decoder performs the following operations:
\begin{eqnarray}
\boldsymbol{Y}_{d} & = & \left[\alpha_{1}\boldsymbol{Y}+\boldsymbol{D}_{1}+\alpha_{1}\boldsymbol{D}_{2}\right]\textrm{ mod }\Lambda_{1}\nonumber \\
 & = & \left[\alpha_{1}\boldsymbol{X}_{1}+\alpha_{1}\boldsymbol{X}_{2}+\alpha_{1}\boldsymbol{Z}+\boldsymbol{D}_{1}+\alpha_{1}\boldsymbol{D}_{2}\right]\textrm{ mod }\Lambda_{1}\nonumber \\
 & = & \left[\boldsymbol{V}_{1}+\alpha_{1}\boldsymbol{V}_{2}+\alpha_{1}\boldsymbol{X}_{1}-\left(\boldsymbol{V}_{1}-\boldsymbol{D}_{1}\right)\right.\nonumber \\
 &  & \left.-\alpha_{1}\mathcal{Q}_{\Lambda_{2}}\left(\boldsymbol{V}_{2}-\boldsymbol{D}_{2}\right)+\alpha_{1}\boldsymbol{Z}\right]\textrm{ mod }\Lambda_{1}\nonumber \\
 & = & \left[\left[\boldsymbol{V}_{1}+\alpha_{1}\boldsymbol{V}_{2}-\alpha_{1}\mathcal{Q}_{\Lambda_{2}}\left(\boldsymbol{V}_{2}-\boldsymbol{D}_{2}\right)\right]\textrm{ mod }\Lambda_{1}\right.\nonumber \\
 &  & \left.+(\alpha_{1}-1)\boldsymbol{X}_{1}+\alpha_{1}\boldsymbol{Z}\right]\textrm{ mod }\Lambda_{1}\label{eq:Effective Noise g>1}\\
 & = & \left[\boldsymbol{T}_{1}+\boldsymbol{Z}_{{\rm eff}}\right]\textrm{ mod }\Lambda_{1},\nonumber 
\end{eqnarray}
where (\ref{eq:Effective Noise g>1}) follows from the distributive
law of the modulo operation. The effective noise is given by
\[
\boldsymbol{Z}_{{\rm eff}}=\left[(\alpha_{1}-1)\boldsymbol{X}_{1}+\alpha_{1}\boldsymbol{Z}\right]\textrm{ mod }\Lambda_{1},
\]
and the sequence to be estimated is given by
\[
\boldsymbol{T}_{1}=\left[\boldsymbol{V}_{1}+\alpha_{1}\boldsymbol{V}_{2}-\alpha_{1}\mathcal{Q}_{\Lambda_{2}}\left(\boldsymbol{V}_{2}-\boldsymbol{D}_{2}\right)\right]\textrm{ mod }\Lambda_{1}.
\]
Due to the dithers, the vectors $\boldsymbol{V}_{1},\boldsymbol{X}_{1}$
are independent, and also independent of $\boldsymbol{Z}$. Therefore,
$\boldsymbol{Z}_{{\rm eff}}$ is independent of $\boldsymbol{V}_{1}$
and $\boldsymbol{V}_{2}$. The decoder attempts to recover $\boldsymbol{T}_{1}$
from $\boldsymbol{Y}_{d}$ instead of recovering $\boldsymbol{V}_{1}$
and $\boldsymbol{V}_{2}$ individually. The method of decoding is
minimum Euclidean distance lattice decoding \cite{Erez_IT_04,Poltyrev_IT_94},
which finds the closest point to $\boldsymbol{Y}_{d}$ in $\Lambda_{c}$.
Thus, the estimate of $\boldsymbol{T}_{1}$ is given by
\[
\hat{\boldsymbol{T}}_{1}=\mathcal{Q}_{\Lambda_{c}}\left(\boldsymbol{Y}_{d}\right),
\]
and the probability of decoding error is given by
\begin{eqnarray*}
P_{e} & = & \textrm{ Pr}\left\{ \hat{\boldsymbol{T}}_{1}\neq\boldsymbol{T}_{1}\right\} =\textrm{ Pr}\left\{ \boldsymbol{Z}_{{\rm eff}}\notin\mathcal{V}_{c}\right\} .
\end{eqnarray*}
\textcolor{black}{As it is shown in \cite{Erez_IT_04} and \cite{Poltyrev_IT_94},
the error probability vanishes as $n\rightarrow\infty$ if
\begin{equation}
\mu=\frac{\left(\mbox{ Vol}\left(\mathcal{V}_{c}\right)\right)^{\frac{2}{n}}}{2\pi e\textrm{Var}\left(\boldsymbol{Z}_{{\rm eff}}^{*}\right)}>1,\label{eq:Condition of Vanishing of ERROR}
\end{equation}
where $\boldsymbol{Z}_{{\rm eff}}^{*}\sim\mathcal{N}\left(0,\textrm{Var}\left(\boldsymbol{Z}_{{\rm eff}}\right)\right)$.
Since $\Lambda_{c}$ is Poltyrev-good, the condition of (\ref{eq:Condition of Vanishing of ERROR})
is satisfied. For calculating rate $R_{1}$, we have:
\begin{eqnarray}
R_{1} & = & \frac{1}{n}\log\left(\frac{\mbox{ Vol}\left(\mathcal{V}_{1}\right)}{\mbox{ Vol}\left(\mathcal{V}_{c}\right)}\right),\nonumber \\
 & = & \frac{1}{2}\log\left(\frac{\sigma^{2}(\Lambda_{1})}{G(\Lambda_{1})\left(\mbox{ Vol}\left(\mathcal{V}_{c}\right)\right)^{\frac{2}{n}}}\right)\nonumber \\
 & \leq & \frac{1}{2}\log\left(\frac{P}{G(\Lambda_{1})2\pi e\textrm{Var}\left(\boldsymbol{Z}_{{\rm eff}}^{*}\right)}\right)\label{eq:Inequality at error condition}\\
 & \leq & \frac{1}{2}\log\left(\frac{{\rm SNR}}{\left(\alpha_{1}-1\right)^{2}{\rm SNR}+\alpha_{1}^{2}}\right)\label{eq:Rogers Good}
\end{eqnarray}
where (\ref{eq:Inequality at error condition}) follows from (\ref{eq:Condition of Vanishing of ERROR}),
and (\ref{eq:Rogers Good}) is based on Rogers goodness of $\Lambda_{1}$.
}Now, for rate $R_{2}$, we have:
\begin{eqnarray}
R_{2} & = & \frac{1}{n}\log\left(\frac{\alpha_{1}^{n}\mbox{ Vol}\left(\mathcal{V}_{2}\right)}{\mbox{ Vol}\left(\mathcal{V}_{c}\right)}\right)\nonumber \\
 & = & \frac{1}{n}\log\left(\frac{\alpha_{1}^{n}\mbox{ Vol}\left(\mathcal{V}_{2}\right)}{\mbox{ Vol}\left(\mathcal{V}_{1}\right)}\right)+\frac{1}{n}\log\left(\frac{\mbox{ Vol}\left(\mathcal{V}_{1}\right)}{\mbox{ Vol}\left(\mathcal{V}_{c}\right)}\right)\nonumber \\
 & = & \frac{1}{2}\log\left(\frac{G(\Lambda_{1})\alpha_{1}^{2}\sigma^{2}\left(\Lambda_{2}\right)}{G(\Lambda_{2})\sigma^{2}\left(\Lambda_{1}\right)}\right)+R_{1}\label{eq:Rogers-good Property}\\
 & = & R_{1}+\frac{1}{2}\log\left(\alpha_{1}^{2}\right)\nonumber \\
 & \leq & \frac{1}{2}\log\left(\frac{\alpha_{1}^{2}{\rm SNR}}{\left(\alpha_{1}-1\right)^{2}{\rm SNR}+\alpha_{1}^{2}}\right),\label{eq:Rate R_2:Case1}
\end{eqnarray}
where (\ref{eq:Rogers-good Property}) follows from the fact that
lattices $\Lambda_{1}$ and $\Lambda_{2}$ are Rogers-good. Thus,
to estimate $\boldsymbol{T}_{1}$ correctly, from (\ref{eq:Rogers Good})
and (\ref{eq:Rate R_2:Case1}), we get the rate-region $\mathcal{R}_{1}\left(\alpha_{1}\right)$,
where
\begin{eqnarray}
\mathcal{R}_{1}\left(\alpha_{1}\right)= & \left\{ \left(R_{1},R_{2}\right):\,\, R_{1}\leq\frac{1}{2}\log\left(\frac{{\rm SNR}}{\left(\alpha_{1}-1\right)^{2}{\rm SNR}+\alpha_{1}^{2}}\right)\right.\nonumber \\
 & \left.R_{2}\leq\frac{1}{2}\log\left(\frac{\alpha_{1}^{2}{\rm SNR}}{\left(\alpha_{1}-1\right)^{2}{\rm SNR}+\alpha_{1}^{2}}\right)\right\} .\label{eq:R_1(alpha_1)}
\end{eqnarray}
Thus, we have proved the following Theorem which is one of the main
contributions of this paper.
\begin{thm}
For the Gaussian MAC shown in Fig. \ref{fig:Fig1}, if any rate pair
$\left(R_{1},R_{2}\right)$ satisfies the rate constraints given in
(\ref{eq:R_1(alpha_1)}), then, there exist sequences of nested lattices
$\Lambda_{1}\subseteq\frac{\Lambda_{2}}{\alpha_{1}}\subseteq\Lambda_{c}$
such that the following linear combination can be recovered: 
\[
\left[\boldsymbol{V}_{1}+\alpha_{1}\boldsymbol{V}_{2}-\alpha_{1}\mathcal{Q}_{\Lambda_{2}}\left(\boldsymbol{V}_{2}-\boldsymbol{D}_{2}\right)\right]\textrm{ mod }\Lambda_{1},
\]
where $0\leq\alpha_{1}\leq1$. 
\end{thm}
Now, by exchanging the role of two encoders in the preceding theorem
and by following the above-mentioned steps, we can show that if 
\begin{eqnarray}
\mathcal{R}_{2}\left(\alpha_{2}\right)= & \left\{ \left(R_{1},R_{2}\right):\,\, R_{1}\leq\frac{1}{2}\log\left(\frac{\alpha_{2}^{2}{\rm SNR}}{\left(\alpha_{2}-1\right)^{2}{\rm SNR}+\alpha_{2}^{2}}\right)\right.\nonumber \\
 & \left.\qquad R_{2}\leq\frac{1}{2}\log\left(\frac{{\rm SNR}}{\left(\alpha_{2}-1\right)^{2}{\rm SNR}+\alpha_{2}^{2}}\right)\right\} ,\quad\nonumber \\
 & 0\leq\alpha_{2}\leq1\label{eq:R_2(alpha_2)}
\end{eqnarray}
then, we can correctly recover the following linear combination at
the destination: 
\[
\boldsymbol{T}_{2}=\left[\boldsymbol{V}_{2}+\alpha_{2}\boldsymbol{V}_{1}-\alpha_{2}\mathcal{Q}_{\Lambda_{1}}\left(\boldsymbol{V}_{1}-\boldsymbol{D}_{1}\right)\right]\textrm{ mod }\Lambda_{2}.
\]
In Fig. \ref{fig:Fig2}, we compare the achievable rate-regions for
estimating these two linear combinations with the outer bound.

\begin{rem}
Note that by replacing $\alpha_{1}$ and $\alpha_{2}$ with $\alpha_{_{{\rm MMSE}}}=\frac{{\rm SNR}}{1+{\rm SNR}}$
in the rate-regions $\mathcal{R}_{1}\left(\alpha_{1}\right)$ and
$\mathcal{R}_{2}\left(\alpha_{2}\right)$, given in (\ref{eq:R_1(alpha_1)})
and (\ref{eq:R_2(alpha_2)}), we can see that the following points
are achievable:
\begin{eqnarray*}
A & = & \left(\frac{1}{2}\log\left(1+{\rm SNR}\right),\frac{1}{2}\log\left(\frac{{\rm SNR}^{2}}{1+{\rm SNR}}\right)\right)\\
B & = & \left(\frac{1}{2}\log\left(\frac{{\rm SNR}^{2}}{1+{\rm SNR}}\right),\frac{1}{2}\log\left(1+{\rm SNR}\right)\right).
\end{eqnarray*}
On the other hand, achieving the sum of codewords over the Gaussian
MAC can be upper bounded by the following rate region:
\begin{eqnarray*}
R_{1} & \leq & \frac{1}{2}\log\left(1+{\rm SNR}\right),\,\,\,\, R_{2}\leq\frac{1}{2}\log\left(1+{\rm SNR}\right).
\end{eqnarray*}
Thus, by comparing this outer bound with our achievable rate region,
we observe that it always coincides with the outer bound for each
linear combination. 
\end{rem}
Achieving linear combination of codewords at the Gaussian MAC has
many applications in network information theory, see e.g. \cite{Wilson_IT_2010_Relay,Nazar_IT_11,Nam_IT_2011}.
In all these papers, achieving $\left[\boldsymbol{V}_{1}+\boldsymbol{V}_{2}\right]\textrm{ mod }\Lambda_{1}$
is studied and it is shown that the following rate region is achievable:
\begin{eqnarray}
R_{i} & \leq & \frac{1}{2}\log\left(\frac{1}{2}+{\rm SNR}\right),\,\,\,\,\,(i=1,2).\label{eq:Ice-Wine Problem R_1}
\end{eqnarray}
\textcolor{black}{By comparing the achievable rate region of the compute-and-forward
scheme, given in (\ref{eq:Ice-Wine Problem R_1}), with the outer
bound, it is clear that the compute-and-forward scheme is not able
to coincide with the outer bound even partially whereas our achievable
rate-region for estimating $\boldsymbol{T}_{1}$ or $\boldsymbol{T}_{2}$
is partially tight.}
\begin{rem}
Here, we compare our proposed scheme for the Ice-Wine problem with
the proposed schemes in \cite{Nazar_IT_11,Wilson_IT_2010_Relay,Nam_IT_2011}.
In these papers, using the fact that each integer linear combination
of lattice points is another lattice point, the given rate-region
in (\ref{eq:Ice-Wine Problem R_1}) is established. As we see, there
is a loss of $\frac{1}{2}$ bit compared with the outer bound. But,
where is the source of this loss?

To achieve $\left[\boldsymbol{V}_{1}+\boldsymbol{V}_{2}\right]\textrm{ mod }\Lambda$,
we are forced to have the following effective noise: 
\[
\boldsymbol{Z}_{{\rm eff}}=\left[\left(\alpha-1\right)\left(\boldsymbol{X}_{1}+\boldsymbol{X}_{2}\right)+\alpha\boldsymbol{Z}\right]\textrm{ mod }\Lambda.
\]
We see that both terms $\boldsymbol{X}_{1}$ and $\boldsymbol{X}_{2}$
are presented in the effective noise. This yields the loss of $\frac{1}{2}$
bit. However, in our proposed scheme, we try to eliminate $\boldsymbol{X}_{2}$
at the effective noise. This helps us to achieve full capacity for
one user but due to the chosen codebooks at the transmitter side,
we cannot achieve the full rate for the other user. 
\end{rem}

\section{\label{sec:The-Gaussian-Two-Way}The Gaussian Two-Way Relay Channel}

\begin{figure}
\begin{centering}
\includegraphics[width=9cm]{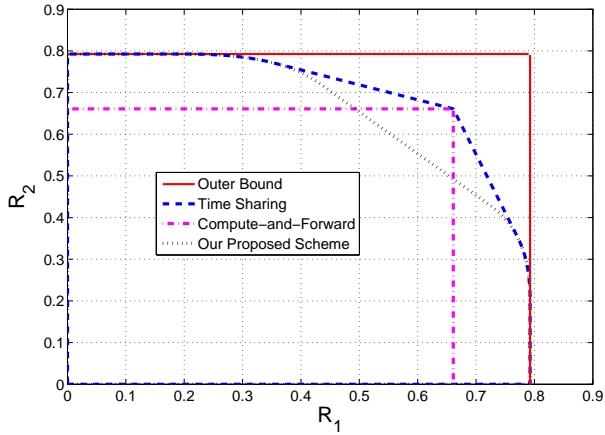}
\par\end{centering}

\caption{\label{fig:Fig3}Achievable rate-regions and rate-region outer bound
for ${\rm SNR}=2$.}

\vspace{-0mm}
 
\end{figure}

\begin{figure}
\begin{centering}
\vspace{-0mm}

\par\end{centering}

\begin{centering}
\includegraphics[width=9cm]{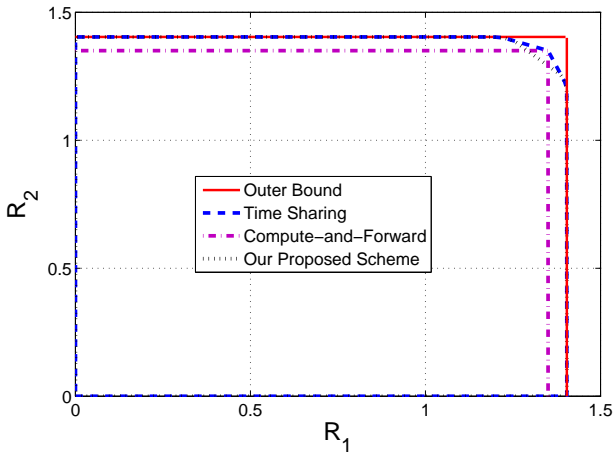}
\par\end{centering}

\caption{\label{fig:Fig4}Achievable rate-regions and rate-region outer bound
for ${\rm SNR}=6$.}
\end{figure}

One can apply the proposed scheme in this paper to the Gaussian Two-Way
Relay Channel (GTWRC) to improve the best rate region of this channel,
provided in \cite{Nam_IT_2010}. The following Theorem provides this
rate-region.
\begin{thm}
For the Gaussian two-way relay channel, if both transmitters transmit
at equal powers, then the following rate-region is achievable:
\begin{equation}
\mathcal{R}=\textrm{cl conv}\left\{ \left(\underset{\alpha_{1}\in\left[0,1\right]}{\bigcup}\mathcal{R}_{1}\left(\alpha_{1}\right)\right)\bigcup\left(\underset{\alpha_{2}\in\left[0,1\right]}{\bigcup}\mathcal{R}_{2}\left(\alpha_{2}\right)\right)\right\} ,\label{eq:Rate Region For the GTWRC}
\end{equation}
where $\mathcal{R}_{1}\left(\alpha_{1}\right)$ and $\mathcal{R}_{2}\left(\alpha_{2}\right)$
are defined in (\ref{eq:R_1(alpha_1)}) and (\ref{eq:R_2(alpha_2)}),
respectively. Also, $\textrm{cl}$ and $\textrm{conv}$ are the closure
and the convex hull operations, respectively.\end{thm}
\begin{IEEEproof}
For this purpose, node $i$ constructs the following sequence and
sends it over the channel:
\[
\boldsymbol{X}_{i}=\left[\boldsymbol{V}_{i}-\boldsymbol{D}_{i}\right]\textrm{ mod }\Lambda_{i},\qquad i=1,2
\]
Then, by our proposed scheme, we can estimate the linear combination,
$\boldsymbol{T}_{1}$ or $\boldsymbol{T}_{2}$, at the relay node.
In the following, without loss of generality, assume that we estimate
$\boldsymbol{T}_{1}$, given as
\[
\boldsymbol{T}_{1}=\left[\boldsymbol{V}_{1}+\alpha_{1}\boldsymbol{V}_{2}-\alpha_{1}\mathcal{Q}_{\Lambda_{2}}\left(\boldsymbol{V}_{2}-\boldsymbol{D}_{2}\right)\right]\textrm{ mod }\Lambda_{1}.
\]
Now, the relay node, using random coding sends $\boldsymbol{T}_{1}$
to both nodes as it is explained in \cite{Nam_IT_2010}. At node 1,
we know $\boldsymbol{V}_{1}$. Thus, we estimate $\boldsymbol{V}_{2}$
as the following:
\begin{eqnarray*}
\widehat{\boldsymbol{V}_{2}} & = & \frac{1}{\alpha_{1}}\left[\boldsymbol{T}_{1}-\boldsymbol{v}_{1}\right]\textrm{ mod }\Lambda_{3}\\
 & = & \frac{1}{\alpha_{1}}\left[\alpha_{1}\boldsymbol{V}_{2}-\alpha_{1}\mathcal{Q}_{\Lambda_{2}}\left(\boldsymbol{V}_{2}-\boldsymbol{D}_{2}\right)\right]\textrm{ mod }\Lambda_{3}\\
 & = & \frac{1}{\alpha_{1}}\left[\alpha_{1}\boldsymbol{V}_{2}\right]\textrm{ mod }\Lambda_{3}=\boldsymbol{V}_{2}.
\end{eqnarray*}
Similarly, node 2 with knowing $\boldsymbol{V}_{2}$, estimates message
of node 1. Thus, we can achieve the rate-region $\mathcal{R}_{1}\left(\alpha_{1}\right)$
for the GTWRC. On the other hand, by finding $\boldsymbol{T}_{2}$
at the relay node, we can see that the rate-region $\mathcal{R}_{2}\left(\alpha_{2}\right)$
is also achievable. Finally, using time-sharing between these two
rate-regions, we get the entire achievable rate-region $\mathcal{R}$
in (\ref{eq:Rate Region For the GTWRC}). 
\end{IEEEproof}
As a numerical example, in Figs. \ref{fig:Fig3} and \ref{fig:Fig4},
we compare the achievable rate-region of our proposed scheme with
that of the compute-and-forward scheme. For comparison, an outer bound
is also provided. As we observe, our proposed scheme can achieve the
outer bound and thus capacity region is partially known. By increasing
SNR, the gap between our proposed scheme with the outer bound is reduced.
In these Figures, we also depict the convex hull of our achievable
rate-region and the achievable rate-region by the compute-and-forward
scheme. To the best of our knowledge, this rate region is the best
rate region for the Gaussian TWRC so far.

\section{\label{sec:Conclusion}Conclusion}

In this paper, we studied the Ice-Wine problem and using nested lattice
codes, we obtained a new achievable rate-region for this problem.
In contrast with the previous obtained achievable rate regions, the
achievable rate-region achieves the outer bound partially for each
user. As we observed, our proposed scheme achieves some rates which
are not achievable by all known schemes to date. Finally, using applying
our proposed scheme to the GTWRC, we showed that the best achievable
rate-region for this open problem can be improved significantly.

\section*{Acknowledgment}

The authors would like to thank the anonymous reviewers for their
valuable comments that have certainly improved the quality of this
paper. The authors also would like to thank Amin Gohari for his helpful
comments.

\bibliographystyle{IEEEtran}
\bibliography{IEEEabrv,ReferencesGhasemiMay2014}

\end{document}